\documentstyle[seceq,twocolumn,epsf]{jpsj}
\newcommand{\rmd}{{\rm d}}

\newcommand{\iu}{{\rm i}}

\def\runtitle{Numerical Study of Impurity Effects on Quasiparticles ...}
\def\runauthor{Yusuke {\sc Kato} and Nobuhiko {\sc Hayashi}}
\title
{Numerical Study of Impurity Effects on Quasiparticles\\
 within S-wave and Chiral P-wave Vortices}
\author
{Yusuke {\sc Kato}\footnote{E-mail: yusuke@phys.c.u-tokyo.ac.jp}
 and
Nobuhiko {\sc Hayashi}$^{1,}$\footnote{E-mail: hayashi@mp.okayama-u.ac.jp}}
\inst
{Department of Basic Science, University of Tokyo, Komaba 3-8-1 Meguro-ku, Tokyo 153-8902\\
and\\ 
$^1$Computer Center, Okayama University, Tsushima-naka 3-1-1 Okayama 700-8530}
\recdate{\hspace{3cm}                                           }
\abst
{The impurity problems within vortex cores of two-dimensional s-wave and chiral p-wave superconductors are studied numerically in the framework of the quasiclassical theory of superconductivity and self-consistent Born approximation under a trial form of the pair potential. The dispersion and impurity scattering rate (the inverse of the relaxation time) of the Andreev bound state localized in vortex cores are deduced from the angular-resoloved local density of states. The energy dependence of the impurity scattering rates depends on the pairing symmetry; particularly, in the chiral p-wave vortex core where chirality and vorticity have opposite sign and hence the total angular momentum is zero, the impurities are ineffective and the scattering rate is vanishingly small. Owing to the cancellation of angular momentum between chirality and vorticity, the chiral p-wave vortex core is similar to locally realized s-wave region and therefore non-magnetic impurity is harmless as a consequence of Anderson's theorem. The results of the present study confirm the previous results of analytical study (J. Phys. Soc. Jpn. {\bf 69} (2000) 3378) in the Born limit.}
\kword{superconductivity, vortex, quasiclassical theory, impurity effect, self-consistent Born approximation, chiral superconductor, flux flow conductivity}
\begin{document}
\sloppy
\maketitle
\section{Introduction}
Vortex cores are often regarded as normal regions with the radius of the coherence length $\xi_0$. This picture is valid only in the dirty superconductors where the mean free path $l$ is much smaller than $\xi_0$. In clean superconductors where $l\gg \xi_0$, quasiparticles experience the Andreev reflections\cite{Andreev} much often than collisions with impurities. The constructive interference at the multiple Andreev reflections leads to the Andreev bound states\cite{Kulik} in the cores; these states are nothing but the Caroli-deGennes-Matricon mode\cite{Caroli}. 
We could expect from the different character in quasiparticles that the flux flow conductivity $\sigma_{\rm f}$\cite{Kopnintext} is different between dirty and clean superconductors. This expectation is not fulfilled in isotropic s-wave superconductors; $\sigma_{\rm f}$ in clean isotropic s-wave superconductors\cite{LO76,BardeenSherman} turns out to be different from that\cite{BardeenStephen} in dirty superconductors only by a factor of $\ln\left(\Delta_\infty/T\right)$ at temperature $T$, where $\Delta_\infty$ is the modulus of the pair-potential in the spatially uniform state. 

However, this is not the whole story. Volovik\cite{Volovik} pointed out that a single impurity in vortex cores of chiral p-wave superconductors~\cite{Sigrist} does not change the spectrum. At low temperatures, the impurity scattering is expected to be the main process of relaxation. Further, the relaxation time governs the conductivity in vortex states\cite{KopninLopatin,Stone}. Therefore, if the impurity scattering rate of quasiparticles in chiral p-wave vortex is completely different from those in isotropic s-wave vortex, we expect different behaviors of flux flow conductivity between chiral p-wave and isotropic s-wave superconductors.

The chiral p-wave superconducting state is specified by the d-vector\cite{SigristUeda,Mineev} ${\mib d}=\hat{\mib z}\left(p_x\pm\iu p_y\right)$~\cite{Sigrist} and expected to be realized in Sr$_2$RuO$_4$~\cite{Maeno}. The chiral p-wave state has two-fold degeneracy. Each phase is specified by the ^^ ^^ chirality", which we define here as the angular momentum of relative motion of Cooper-pairs; the phase with ${\mib d}=\hat{\mib z}\left[p_x+\iu p_y\right] (\hat{\mib z}\left[p_x-\iu p_y\right])$ has $+1$ $(-1)$ chirality. In the absence of external magnetic field, two homogeneous phases with definite chirality are energetically equal\cite{Sigrist} and hence it depends on the history what phase is realized among the following possibilities: spatially uniform phase with positive or negative chirality or the domain structure consisting of phases with different chiralities. 

\begin{figure}
\vspace{1cm}
\begin{center}
\epsfysize=70pt
\epsfbox{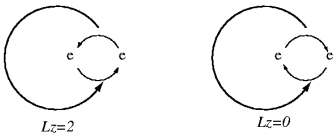}
\end{center}
\caption{Schematic description of chiral p-wave vortex. Cooper-pairs of chiral p-wave vortex states have $\pm 1$ angular momentum ( chirality ) in the internal motion. In the presence of vortex, the center-of-mass motion of Cooper-pairs has the angular momentum (vorticity) around a vortex center. The total angular momentum is given by the sum of chirality and vorticity}
\label{fig:1}
\end{figure}

The magnetic field lifts the degeneracy of the two homogeneous phases\cite{HeebAgterberg}. If the external field induces vortices with positive vorticity, the phase with negative chirality becomes stable and that with positive chirality becomes metastable. When the system is axisymmetric, the center-of-mass motion of Cooper-pairs acquires angular momentum (vorticity) with respect to the vortex center. Total angular momentum is given by the sum of vorticity and chirality. In terms of the total angular momentum around vortex center, the phase with $L_z=2$ is metastable and the phase with $L_z=0$ is stable. These two phases are schematically described in Fig.~\ref{fig:1}.

Although the degeneracy is lifted in vortex state, Ginzburg-Landau theory gives small difference of energy between the two states\cite{HeebAgterberg}. Therefore we expect that the phase $L_z=2$ is relevant to experiments through the domain phase. We expect that the phase with $L_z=0$ is realized in field-cooling experiments and domain structure consisting of the phase with $L_z=0$ and $L_z=2$ is realized in zero-field-cooling experiments . We consider, in the following part, single vortex $L_z=0$ and $L_z=2$. We also consider isotropic s-wave vortex, which has $L_z=1$, as a reference. 

By examining the impurity scattering rates, or the inverse of impurity relaxation time, we can find the evidence that quasiparticles in vortex cores are different from that in normal states.
The energy dependence of $\Gamma$ (normalized by the normal state impurity scattering rate $\Gamma_{\rm n}$) of quasiparticles (Andreev bound states) localized in vortex cores is given by
\begin{equation}
\Gamma/\Gamma_{\rm n}=\left\{
\begin{array}{cl}
c_1\ln\left(\Delta_\infty/E\right),&\mbox{for $L_z=1$ (s-wave)\cite{KopninLopatin}}\\
c_2,&\mbox{for $L_z=2$ (chiral p-wave)\cite{Kato}}\\
0,&\mbox{for $L_z=0$ (chiral p-wave)\cite{Kato}}
\end{array}
\right.
\label{Gamma}
\end{equation}
within the non-selfconsistent Born approximation\cite{KopninLopatin} or the self-consistent Born approximation with a restriction on the energy range\cite{Kato,Kopnin99}. Here $c_1$ and $c_2$ are independent of energy. These two are of the order of ${\cal O}(1/\ln\left(\Delta_\infty/T\right))$ when the vortex shrinking (the Kramer-Pesch effect\cite{KramerPesch}) occurs; otherwise they are of the order of unity. The phase of the pair-potential depends on the pairing symmetry. Further the sensitivity on the phase of the pair-potential is characteristic of the Andreev bound states. Therefore, the impurity scattering rate of the Andreev bound states depends on the pairing symmetry. From (\ref{Gamma}), we can see that in s-wave vortex, the bound states with smaller energy are scattered more strongly by impurities. In chiral p-wave vortex with $L_z=2$, the bound states with various energy are scattered equally. In chiral p-wave vortex with $L_z=0$, the bound states are not subject to the impurity scattering; this point is worthy of attention. The cancellation of angular momentum for a chiral p-wave vortex makes the physics in vortex cores similar to that in s-wave superconductors in zero magnetic field. We know that non-magnetic impurities do not affect superconducting properties in the s-wave superconductors in the absence of magnetic field (the Anderson theorem\cite{Anderson}). In ref.~\citen{KatoHayashi}, the authors have interpreted the vanishing $\Gamma$ in chiral p-wave vortices with $L_z=0$ as a consequence of {\it novel applicability of the Anderson theorem in those vortex cores}. 

The difference of the energy dependence of $\Gamma$ (\ref{Gamma}) leads to different temperature dependence of flux flow conductivity for the three pairing symmetries; with the basis of (\ref{Gamma}) and consideration of the vortex shrinking effect (the Kramer-Pesch effect), the flux flow conductivity $\sigma_{\rm f}$ normalized by the normal state conductivity $\sigma_{\rm n}$ is expected to be\cite{Kato,Kopnintext}
\begin{eqnarray}
& &\sigma_{\rm f}/\sigma_{\rm n}\nonumber\\
& &\left\{
\begin{array}{cl}
\sim\ln\left(\Delta_\infty/T\right)H_{\rm c2}/B,&\mbox{for $L_z=1$ (s-wave)\cite{LO76}}\\
\sim\left[\ln\left(\Delta_\infty/T\right)\right]^2 H_{\rm c2}/B,&\mbox{for $L_z=2$ (chiral p-wave)\cite{Kato}}\\
\gg{\cal O}(\Delta_\infty/T)H_{\rm c2}/B,&\mbox{for $L_z=0$ (chiral p-wave)\cite{Kato}}
\end{array}
\right.
\label{sigma}
\end{eqnarray}
at temperature $T$ and under the magnetic field $B$ for clean superconductors with the upper critical field $H_{\rm c2}$. 
Thus, it is of great importance to confirm the result (\ref{Gamma}) on the energy dependence of impurity scattering rate in order to discuss the flux flow (\ref{sigma}).
 
\begin{figure}
\begin{center}
\epsfysize=150pt
\epsfbox{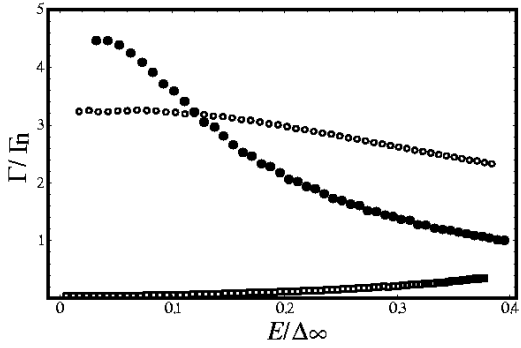}
%\epsfbox{gammaene_born_lz012.eps}
\end{center}
\caption{The impurity scattering rate $\Gamma$ as a function of energy of the Andreev bound states. Solid circles represent the results for s-wave vortex with $\Gamma_{\rm n}/\Delta_\infty=0.01$, open circles for chiral p-wave vortex with $L_z=2$ and $\Gamma_{\rm n}/\Delta_\infty=0.01$, and open squares for chiral p-wave vortex with $L_z=0$ and $\Gamma_{\rm n}/\Delta_\infty=0.1$}
\label{fig: gammaene}
\end{figure}

In this paper, we examine the results (\ref{Gamma}) by numerical calculations. The scheme we adopt is the quasiclassical theory of superconductivity\cite{Eilenberger,LO68, Eliashberg,SereneRainer} and the {\it self-consistent} Born approximation. Before going into the description of model, methods and other details, we present main result in this paper. 
Figure~\ref{fig: gammaene} shows the impurity scattering rate as a function of energy of Andreev bound states. The impurity scattering rate within an s-wave vortex with $\Gamma_{\rm n}/\Delta_\infty=0.01$ shown by solid circles reveals the logarithmic energy dependence. For chiral p-wave vortex with $L_z=2$ and $\Gamma_{\rm n}/\Delta_\infty=0.01$ (open circles), the impurity scattering rate is less dependent on energy and particularly becomes independent of energy for $E/\Delta_\infty<0.1$. For chiral p-wave vortex with $L_z=0$ and $\Gamma_{\rm n}/\Delta_\infty=0.1$ (solid squares), the impurity scattering rate is extremely small. 
From these observations, we confirm the previous result\cite{KopninLopatin,Kato} (\ref{Gamma}). Accordingly, the result on the flux flow conductivity (\ref{sigma}) becomes more convincing. 
 
The rest of this paper is organized as follows.
In the next section, we first summarize the formulation of quasiclassical theory of superconductivity\cite{Eilenberger,LO68, Eliashberg,SereneRainer} and self-consistent Born approximation, next explain the way how we can investigate the bound state in the presence of impurities within the green function formalism and third give a brief description on numerics. In section III, we present numerical results on the angular resolved local density of states for the three pairing symmetries. From those results, we can obtain the impurity scattering rate shown in Fig.~\ref{fig: gammaene}. In section IV, we discuss the significance of our findings, in comparison with earlier studies. In section V, we summarize our conclusions.  
\section{Model and Method }
\subsection{quasiclassical theory of superconductivity}
We study the impurity effects in vortex cores in two-dimensional superconductors with s-wave and chiral p-wave pairing symmetries. For simplicity, we consider the systems in the type II limit with the circular symmetric Fermi surface. In the quasiclassical theory of superconductivity, s-wave and chiral p-wave superconductors can be studied on equal-footing. In what follows, we adhere to the notations in ref.~\citen{Kato}. What we calculate and discuss in this paper is the retarded part of the quasiclassical green function in the equilibrium case
\begin{equation}
\hat g(\epsilon, {\mib r},\hat {\mib p})=\hat g(\epsilon,{\mib r},\alpha)
=\left(
\begin{array}{rc}
g&f\\
-\tilde f&-g
\end{array}\right),
\label{greenfunction}
\end{equation}
which is a $2\times 2$ matrix in particle-hole space and is a function of frequency $\epsilon$, the direction $\hat{\mib p}=(\cos\alpha,\sin\alpha)$ of momentum ${\mib p}=p_{\rm F}\hat{\mib p}$ and the point ${\mib r}=r(\cos \phi,\sin \phi)$ in real space. The quasiclassical green function satisfies the normalization condition 
\begin{equation}
\hat g^2=-\pi^2 \hat 1\quad (\hat 1 \mbox{; the 2$\times$ 2 unit matrix}), 
\label{normalization}
\end{equation}
and follows the equation of motion (the Eilenberger equation)\cite{Eilenberger}
\begin{equation}
-\iu  {\mib v}\cdot{\mib\nabla} \hat g=\left[\left(\epsilon+\iu \delta\right) \hat\tau_3-\hat \Delta-\hat\Sigma ,\hat g\right]. 
\label{qclequation}
\end{equation}
Here ${\mib v}=v\hat {\mib p}$ is the Fermi velocity, $\delta$ is a positive infinitesimal and $\hat \tau_3$ is a Pauli matrix
\begin{equation}
\left(
\begin{array}{rc}
1&0\\
0&-1
\end{array}\right). 
\label{pauli3}
\end{equation}
The matrix $\hat \Delta=\hat \Delta({\mib r},\hat{\mib p})$ in eq. (\ref{qclequation}) is given by
\begin{equation}
\left(
\begin{array}{rc}
0&\Delta({\mib r},\hat{\mib p})\\
-\Delta^*({\mib r},\hat{\mib p})&0
\end{array}\right), 
\label{deltahat}
\end{equation}
in terms of the pair-potential $\Delta({\mib r},\hat{\mib p})$. The $\hat{\mib p}$ (or equivalently $\alpha$) dependence of the pair-potential is determined by the pairing symmetry. The self-energy $\hat \Sigma=\hat\Sigma(\epsilon, {\mib r})$ in the right hand side of (\ref{qclequation}) is given by that in the self-consistent Born approximation
\begin{equation}
\Gamma_{\rm n}\langle\hat g(\epsilon,{\mib r},\alpha)\rangle,
\label{selfenergy}
\end{equation}
where $\Gamma_{\rm n}$ is the impurity scattering rate in the normal state and $\langle\cdots\rangle=\int\cdots\rmd \alpha/(2\pi)$ denotes the average over the Fermi surface. 

We assume that the pair-potential $\Delta({\mib r},\hat{\mib p})$ has the following form:
\begin{equation}
\Delta({\mib r},\hat{\mib p})=\Delta_0(r){\rm e}^{\iu \left(\phi -\alpha\right)+\iu L_z \alpha}. 
\label{Deltaform}
\end{equation}
Here $\Delta_0(r)$ is a monotonically increasing function satisfying the following conditions: $\Delta_0(r=0)=0$ and $\lim_{r\rightarrow \infty}\Delta_0(r)=\Delta_\infty$. The expression (\ref{Deltaform}) describes isolated vortices with vorticity $+1$ at ${\mib r}=0$. Under a given form of the pair-potential, $\hat g$ and $\hat \Sigma$ are self-consistently determined in numerical calculations. 
For a given $\hat p$ or $\alpha$, a rotated frame in real space
\begin{equation}
{\mib r}=s\hat {\mib p} +b \hat{\mib z}\times \hat {\mib p}
\end{equation}
with $s=\cos\left(\phi-\alpha\right)$ and $b=r\sin\left(\phi-\alpha\right)$ is suitable to study eq.~(\ref{qclequation}). The left hand side of eq.~(\ref{qclequation}) can be rewritten as $-\iu v\partial \hat g/\partial s$ and the equation yields a one-dimensional problem on the line (the quasiclassical trajectory) with constant $b$. Now $s$ and $b$ turn out to be, respectively, the coordinate along the quasiclassical trajectory and the impact parameter with respect to the vortex center. 
\begin{figure}
\begin{center}
\epsfysize=200pt
\epsfbox{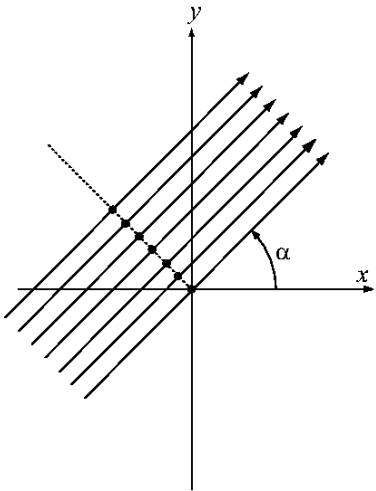}
\end{center}
\caption{Quasiclassical trajectories, on which the Eilenberger equation is to be solved, are shown for a given direction $\alpha$ of momentum of quasiparticles. Solid circles represent the points where we calculate the angular-resolved local density of states. }
\label{fig: coordinate}
\end{figure}
\subsection{Andreev bound states and angular resolved local density of states}
For energy much lower than the modulus of the pair-potential $\Delta_\infty$ in the bulk, most spectral weight of $\hat g$ is exhausted by the contribution with the impact parameter $b$ whose modulus is much smaller than the coherence length $\xi_0=v/(\pi \Delta_\infty)$. In the absence of impurities (and hence the self-energy in eq.~(\ref{qclequation})), the quasiclassical green function for $|\epsilon|\ll \Delta_\infty$ and $|b|\ll \xi_0$, is expected to have the form of\cite{KramerPesch,EschrigThesis,Kato,Kopnintext} 
\begin{equation}
\hat g(\epsilon, {\mib r},\alpha)=\frac{\pi v{\rm exp}\left[-u(s)\right]\hat M(\alpha)}{C\left(\epsilon -E_{\rm p}(b)+\iu \delta\right)}+\mbox{(regular part)},
\label{gapproximate}
\end{equation}
as a function of $\epsilon$, $s$, $b$ and $\alpha$. 
Here the function $u(s)$ in the numerator is defined by
\begin{equation}
u(s)=\frac{2}{v}\int_0^{\left|s\right|}\rmd s' \Delta_0(s') 
\label{u}
\end{equation}
and the matrix $\hat M(\alpha)$ is defined by
\begin{equation}
\hat M(\alpha)\equiv \left(
\begin{array}{ccc}
1&-\iu {\rm e}^{\iu L_z\alpha}\\
-\iu {\rm e}^{-\iu L_z\alpha}&-1
\end{array}\right). 
\label{Malpha}
\end{equation}
The constant $C$ in the denominator denotes the normalization factor 
\begin{equation}
C=\int_0^\infty\rmd s\exp\left[-u(s)\right]\sim \xi_0
\end{equation}
and $E_{\rm p}(b)$ is given by
\begin{equation}
E_{\rm p}(b)=\frac{b}{C}\int_0^\infty\rmd s\frac{\Delta_0(s)}{s}{\rm e}^{-u(s)}.\label{Epb}
\end{equation}
The subscript p stands for pure superconductors. 
The second term in the right hand side of (\ref{gapproximate}) represents negligible contributions from the scattering states of quasiparticles. 

From the above results, we see that $N(\epsilon,\alpha,s,b)/N_0=-{\rm Im}g(\epsilon, {\mib r},\alpha)$, which is ^^ ^^ the angular-resolved local density of states"\cite{Klein,RainerSaulsWaxman,EschrigThesis} normalized by the density of states $N_0$ in the normal state, has the following expression:
\begin{equation}
N(\epsilon,\alpha,s,b)/N_0=\pi^2 v C^{-1}{\rm e}^{-u(s)}\delta(\epsilon-E_{\rm p}(b)).
\label{eq: arldos}
\end{equation}
This expression (\ref{eq: arldos}) has a sharp peak along a trajectory shown in Fig.~\ref{fig: coordinate}; on the trajectory, the relation $\epsilon=E_{\rm p}(b)$ is satisfied for a given $\alpha$. Numerical results on the angular-resolved local density of states were firstly presented by Klein\cite{Klein} for vortex lattice of an s-wave superconductor without impurities. For isolated s-wave vortex with impurities, the angular resoloved local density of states were presented in ref.~\citen{EschrigThesis}. 

In the presence of sufficiently dilute impurities, the expression (\ref{gapproximate}) is expected to be modified as\cite{Kato,Kopnin99}
\begin{equation}
\frac{\pi v{\rm exp}\left[-u(s)\right]\hat M(\alpha)}{C\left(\epsilon -E(b)+\iu \Gamma\right)}+\mbox{(regular part)},
\label{gapproximate2}
\end{equation}
and accordingly the expression (\ref{eq: arldos}) turns into
\begin{equation}
N(\epsilon,\alpha,s,b)/N_0=\frac{\pi^2 v C^{-1}{\rm e}^{-u(s)}\Gamma}{\left[(\epsilon-E(b))^2+\Gamma^2\right]}, 
\label{arldos2}
\end{equation}
where $E(b)$ and $\Gamma$ are certain functions of $b$ and are independent of $s$ for a given $\alpha$. 
These results are suggestive of the way to investigate the energy of Andreev bound states and the impurity scattering rate in impure but clean superconductors; if $N(\epsilon,\alpha,s,b)/N_0$ has a single peak as a function of $\epsilon$, the peak position gives the energy $E(b)$ of Andreev bound states for a given impact parameter $b$. The width $\Gamma$ of the single peak, on the other hand, gives the impurity scattering rate for the Andreev bound states. In the next section, we will present numerical results on $N(\epsilon,\alpha,s,b)/N_0$ as functions of $\epsilon$ for a certain fixed value of $\alpha$, $s(=0)$ and various values of $b$, or equivalently for the points shown by solid circles in Fig.~\ref{fig: coordinate}. Furthermore, owing to the axisymmetry, $N(\epsilon,\alpha,s=0,b)/N_0$ is independent of $\alpha$. We can thus set $\alpha=0$  without loss of generality.

\subsection{Numerical details}
We take the profile of the pair-potential as 
\begin{equation}
\Delta_0(r)=\Delta_\infty\tanh(r/\xi_0).
\label{tanh}
\end{equation}
The Eilenberger equation (\ref{qclequation}) is solved numerically by the fourth-order Runge-Kutta method after transforming to the Riccati formulation
\cite{Nagato1,Higashitani,Nagato2,SchopohlMaki,Schopohl,EschrigThesis} of (\ref{qclequation}).
The cut-off radius, from which the Runge-Kutta program starts, is taken as $3\xi_0$. In the Runge-Kutta method, the choice of the increment step being  $3\times 10^{-3}\xi_0$ along the quasiclassical trajectory yields the bounds of relative error $10^{-4}$. The $\alpha$-integration on the 2D Fermi surface is replaced by the sum over the points $\alpha_j=2\pi j/N_\alpha$ with $j\in [1,N_\alpha]$. The choice of $N_\alpha=3000$ for $\Gamma_{\rm n}=0.1\Delta_\infty$ is sufficient to obtain the upper bound of the error of the self-energy $\delta\hat\Sigma$ 
\begin{equation}
{\rm Max}(\left|\delta\Sigma_{11}\right|,\left|\delta\Sigma_{12}\right|,\left|\delta\Sigma_{21}\right|)<10^{-3}\Gamma_{\rm n}.
\label{selfaccuracy}
\end{equation}
 In our calculation, the self-consistent condition on the self-energy is satisfied within the accuracy (\ref{selfaccuracy}).
\section{Results}\label{sec: Results}
\subsection{S-wave vortex}
\begin{figure}
\begin{center}
\epsfysize=300pt
\epsfbox{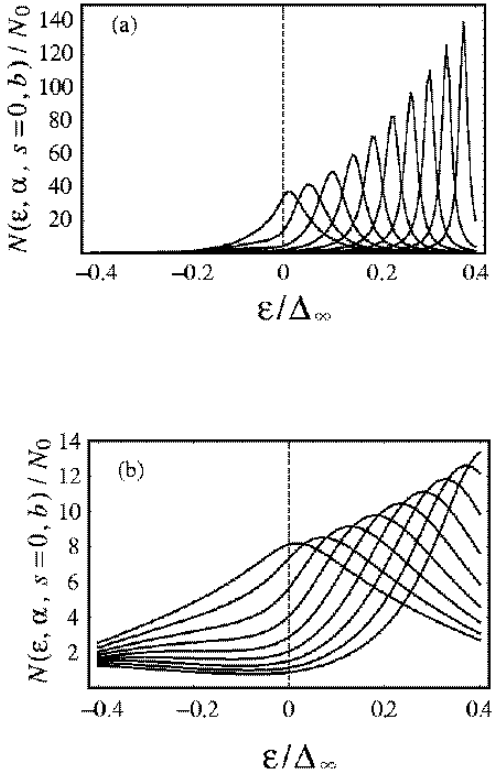}
%\epsfysize=100pt
%\epsfbox{peak_born_s_tau01_nimp80.eps}
\end{center}
\caption{The angular resolved local density of states \\
$N(\epsilon, \alpha=0,s=0,b)/N_0$ of s-wave superconductors for $b/\xi_0=0.01$ and $0.05n$ with $n=1,2,\cdots$ for (a) $\Gamma_{\rm n}=0.01\Delta_\infty$ and (b) $\Gamma_{\rm n}=0.1\Delta_\infty$. For both cases, the smaller the impact parameter is, the width of the peak becomes broader. }
\label{fig: spectralswave}
\end{figure}
\begin{figure}
\begin{center}
\epsfysize=150pt
\epsfbox{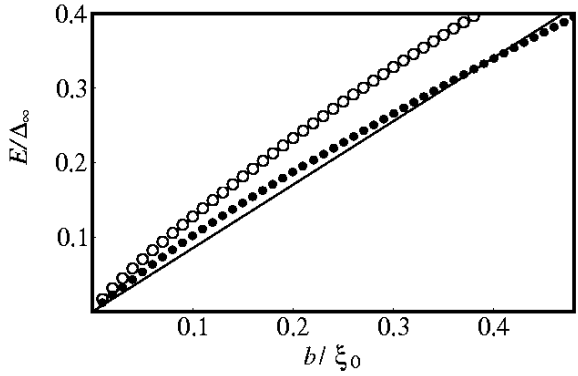}
\end{center}
\caption{The bound state energy as a function of the impact parameter for s-wave vortex. Solid circles represent the results for $\Gamma_{\rm n}=0.01\Delta_\infty$ and open circles $\Gamma_{\rm n}=0.1\Delta_\infty$. The line represents analytical result $E_{\rm p}(b)=0.852557\Delta_\infty b/\xi_0$, which is obtained by substituting (\ref{tanh}) into (\ref{Epb}) for pure superconductors. }
\label{fig: beneswave}
\end{figure}

Figure~\ref{fig: spectralswave} shows $\epsilon$ dependence of the angular-resolved local density of states of the quasiparticles with momentum direction $\alpha=0$ and at the positions (shown by solid circles in Fig.~\ref{fig: coordinate}) of $s=0$ and  $b/\xi_0=0.01$ and $0.05n$ with $n=1,2,\cdots$ (from left to right). The upper panel (a) represents the results for $\Gamma_{\rm n}=0.01\Delta_\infty$ and the lower panel (b) for $\Gamma_{\rm n}=0.1\Delta_\infty$. Each curve for a given impact parameter $b$ has a single peak; This fact shows that the picture of the bound states still works even in the presence of impurities. First we discuss the peak position. The smaller the impact parameter $b$ is, the peak position shifts to the lower energy. Figure~\ref{fig: beneswave} shows that the energy of bound states for a given impact parameter for  $\Gamma_{\rm n}=0.01\Delta_\infty$ (solid circle) and $\Gamma_{\rm n}=0.1\Delta_\infty$ (open circle). The line represents analytical result $E_{\rm p}(b)=0.852557\Delta_\infty b/\xi_0$, which is obtained by substituting (\ref{tanh}) into (\ref{Epb}) for pure superconductors. The bound state energy is nearly linear in the impact parameter. This bound state energy for $\Gamma_{\rm n}=0.01\Delta_\infty$ is almost same as that of pure case. The bound state energy for $\Gamma_{\rm n}=0.1\Delta_\infty$ is enhanced compared to the pure case, owing to the self-energy effect. 

Next we discuss the width of each curve in fig.~\ref{fig: spectralswave}. In the figures, the smaller the impact parameter is, the peak becomes broader. This means that the Andreev bound states with smaller energy are scattered more strongly by impurities. This observation agrees qualitatively with the results of non-self consistent calculations mentioned in the Introduction. For $\Gamma_{\rm n}=0.01\Delta_\infty$, each curve fits well to the form of the Lorentzian $1/\left[\left(\epsilon-E\right)^2+\Gamma^2\right]$ up to a constant factor. From this fitting, we obtain the impurity scattering rate as a function of bound state energy, which has been already shown in Fig.~\ref{fig: gammaene}. For $\Gamma_{\rm n}=0.1\Delta_\infty$, on the other hand, each curve does not necessarily fit to the Lorentzian; a comparison of the impurity scattering rate calculated from the curvature near the peak position and that from the full width of the half-height shows appreciable difference, particularly for small impact parameters. Therefore, we leave the discussion for $\Gamma_{\rm n}=0.1\Delta_\infty$ in a qualitative level. 

\subsection{Chiral p-wave vortices with $L_z=2$}
\begin{figure}
\begin{center}
\epsfysize=200pt
\epsfbox{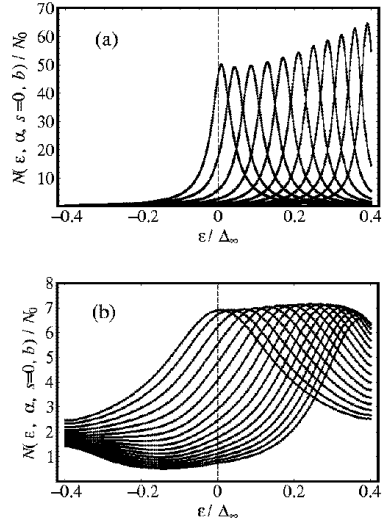}
\end{center}
\caption{The angular resolved local density of states \\
$N(\epsilon, \alpha=0,s=0,b)$ of chiral p-wave superconductors with $L_z=2$ for $b/\xi_0=0.01$ and $0.05n$ with $n=1,2,\cdots$ for (a) $\Gamma_{\rm n}=0.01\Delta_\infty$ and (b) $\Gamma_{\rm n}=0.1\Delta_\infty$. For both cases, the width of all the curves is almost same, in contrast to the s-wave. }
\label{fig: spectrallz2}
\end{figure}

\begin{figure}
\begin{center}
\epsfysize=150pt
\epsfbox{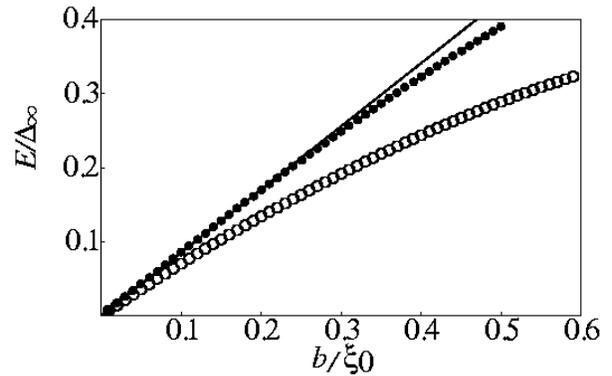}
\end{center}
\caption{The bound state energy as a function of the impact parameter for chiral p-wave vortex. Solid circles represent the results for $\Gamma_{\rm n}=0.01\Delta_\infty$ and open circles $\Gamma_{\rm n}=0.1\Delta_\infty$. The line represents analytical result $E_{\rm p}(b)=0.852557\Delta_\infty b/\xi_0$, which is obtained by substituting (\ref{tanh}) into (\ref{Epb}) for pure superconductors. }
\label{fig: benelz2}
\end{figure}

Figure~\ref{fig: spectrallz2} shows $\epsilon$ dependence of the angular-resolved local density of states for chiral p-wave vortices with $L_z=2$ for (a) $\Gamma_{\rm n}=0.01\Delta_\infty$ and (b) $\Gamma_{\rm n}=0.1\Delta_\infty$. The parameters such as $\alpha$, $s$ and $b$ are taken as the same as that in the s-wave case. 
In figures, more right curve corresponds to larger impact parameter.
Single peak for each impact parameter suggests the existence of a bound state.  The bound state energy obtained from the peak position is shown in fig.~\ref{fig: benelz2}. 
The bound state energy for $\Gamma_{\rm n}=0.01\Delta_\infty$ shown by solid circles is almost same as that of pure superconductors shown by the solid line. On the other hand, the bound state energy $\Gamma_{\rm n}=0.1\Delta_\infty$ shown by open circles is suppressed compared to the pure case. This tendency of the renormalization effect is opposite to that of s-wave case, where the self-energy renormalization yields the enhancement of the bound state energy.  

Now we turn to the width of the peak. The shapes of the curves are almost same for small impact parameters. This means that the impurity scattering rate becomes independent of the impact parameter $b$ for small $b$ or equivalently the bound state energy for small energy. This is confirmed quantitatively; for $\Gamma_{\rm n}/\Delta_\infty=0.01$, all curves fit the Lorentian. The deduced impurity scattering rate has been shown in fig.~\ref{fig: gammaene}. Our results confirm the previous results (\ref{Gamma}) for chiral p-wave case with $L_z=2$.
\subsection{Chiral p-wave vortices with $L_z=0$}
\begin{figure}
\begin{center}
\epsfysize=140pt
%\epsfbox{peak_born_lz0_tau01.eps}
\epsfbox{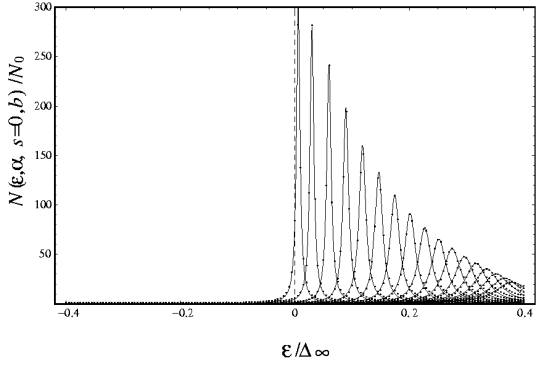}
\end{center}
\caption{The angular resolved local density of states \\
$N(\epsilon, \alpha=0,s=0,b)$ of chiral p-wave superconductors with $L_z=0$ for $b/\xi_0=0.01$ and $0.05n$ with $n=1,2,\cdots$ for $\Gamma_{\rm n}=0.1\Delta_\infty$. Peak width is much smaller than those in s-wave vortex(Fig.\ref{fig: spectralswave}b) and chiral p-wave vortex with $L_z=2$ (Fig.\ref{fig: spectrallz2}b) with the same value of $\Gamma_{\rm n}$.} 
\label{fig: spectrallz0}
\end{figure}

\begin{figure}
\begin{center}
\epsfysize=150pt
%\epsfbox{bene_born_lz0.eps}
\epsfbox{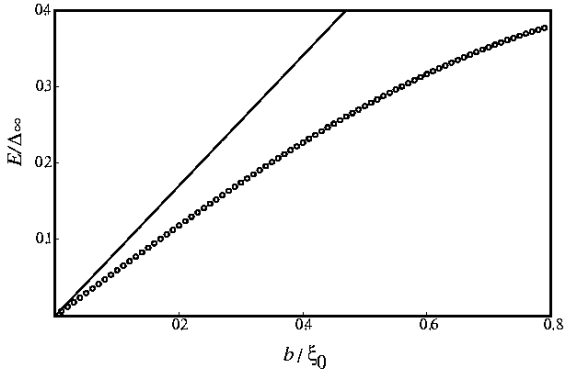}
\end{center}
\caption{The bound state energy as a function of the impact parameter for chiral p-wave vortex. Open circles represent the results for $\Gamma_{\rm n}=0.1\Delta_\infty$. The solid line represents analytical result $E_{\rm p}(b)=0.852557\Delta_\infty b/\xi_0$, which is obtained by substituting (\ref{tanh}) into (\ref{Epb}) for pure superconductors. }
\label{fig: benelz0}
\end{figure}

Figure~\ref{fig: spectrallz0} shows $\epsilon$ dependence of the angular-resolved local density of states for chiral p-wave vortices with $L_z=0$ for $\Gamma_{\rm n}=0.1\Delta_\infty$ (Numerical calculation for $\Gamma_{\rm n}=0.01\Delta_\infty$ turns out to be prohibitive). The parameters such as $\alpha$, $s$ and $b$ are taken as the same as that in the s-wave case. In figures, more right curve corresponds to larger impact parameter. Open circles in fig.~\ref{fig: benelz0} represent the bound state energy obtained from the peak position. The solid line shows the bound state energy in pure superconductors. 
The self-energy leads to the suppression of the bound state energy, contrary to the s-wave case but similarly to the chiral p-wave vortex with $L_z=2$. 

Now we turn to the width of the peak, which is obviously much smaller than those for s-wave (Fig.~\ref{fig: spectralswave}b) and chiral p-wave with $L_z=2$ (Fig.~\ref{fig: spectrallz2}b). The Lorentian-fit works well for all curves and the impurity scattering rate $\Gamma$ is deduced, as shown in Fig.~\ref{fig: gammaene}. Finite but extremely small impurity scattering rate is consistent with the previous results (\ref{Gamma}) for chiral p-wave with $L_z=0$. 
\section{Discussion}

Here we recall earlier studies on the impurity problem of chiral p-wave vortex and make clear the importance of our findings. Volovik\cite{Volovik} studied energy spectrum of quasiparticles bound to vortex cores. His method is a combination of the Andreev approximation of Bogoliubov-deGennes (BdG) equation and Bohr-Sommerfeld semiclassical quantization. He concluded that a single impurity does not change the spectrum of chiral p-wave vortex with $L_z=2$ {\it and} $L_z=0$; it is crucial, in his theory, whether the chirality is odd or even. However, this statement is subject to an counterexample. Matsumoto and Sigrist\cite{M&S} found in numerical calculation of BdG equation of chiral p-wave vortex that the single impurity {\it at the vortex center} shifts the energy spectrum of Caroli-deGennes-Matricon mode for $L_z=2$ and does not for $L_z=0$. Very recently, Miyazu\cite{Miyazu} studied analytically the effect of single impurity at a generic position inside cores of chiral p-wave vortex and found that the impurity is harmless for $L_z=0$ and effective for $L_z=2$. 

The above results are on single impurity in chiral p-wave vortex. However, in realistic situations there are many impurities. Further, there are no explicit implications on physical quantities accessible in experiments. Our previous results\cite{Kato} in (\ref{Gamma}) for chiral p-wave vortex have addressed these issues. The results in the present paper further confirm that the results (\ref{Gamma}) are correct {\it even in the self-consistent treatment of impurity self-energy}. Accordingly we expect that the prediction on the flux flow conductivity (\ref{sigma}) comes true. The measurements of flux flow conductivity of chiral p-wave superconductors (e.g. Sr$_2$RuO$_4$) are highly desirable. 
The absence of impurity effects in vortex cores has another implication; the absence of impurity effects makes the Kramer-Pesch effect much more accessible in experiments of chiral p-wave superconductors than any other superconductors\cite{KatoHayashi}. This can be, in principle, examined experimentally, e.g. by Muon spin relaxation measurements\cite{Sonier}. Further, the cancellation of angular momentum in chiral p-wave vortex yields novel quantum effects on the pinning problem due to impurities\cite{HayashiKato}. This phenomenon can be also tested in the measurements of $M$-$H$ curves in multiple cycles. 
\section{Conclusion}\label{sec:Conclusion}

We studied the impurity effects numerically on quasiparticles localized in vortex cores of s-wave and chiral p-wave superconductors under the scheme of quasiclassical theory of superconductivity and self-consistent Born approximation. From angular-resolved local density of states, we calculated the bound state energy and impurity scattering rates. In s-wave case, the bound states with smaller energy are scattered more strongly by impurities. On the other hand, in the case of the chiral p-wave vortex with $L_z=2$, the impurity scattering rate is independent of energy of the bound state. An intriguing thing occurs in the case of chiral p-wave vortex with $L_z=0$; the impurity scattering rate is almost vanishing. 
\acknowledgement
The authors thank M.~Sigrist, N.~Schopohl, M.~Zirnbauer, M.~Matsumoto, A.~Tanaka, Y.~Matsuda, K.~Izawa, T.~Tamegai, Y.~Maeno for their useful discussions. 
Y. K also thanks D.~Yoshioka and N.~Shibata 
for their help on a computer facility. 
This work is supported by Grant-in-Aid for Scientific Research on Priority Areas (A) of ^^ ^^ Novel Quantum Phenomena in Transition Metal Oxides" (12046225) from the Ministry of Education, Science, Sports and Culture and Grant-in-Aid for Encouragement of Young Scientists from Japan Society for the Promotion of Science (12740203). 

\end{document}